\documentclass[conference]{IEEEtran}
\IEEEoverridecommandlockouts
\usepackage{cite}
\usepackage{amsmath,amssymb,amsfonts}
\usepackage{algorithmic}
\usepackage{graphicx}
\usepackage{textcomp}
\usepackage{multirow}
\usepackage{xcolor}
\usepackage[numbers]{natbib}

 \usepackage{stfloats}
\def\BibTeX{{\rm B\kern-.05em{\sc i\kern-.025em b}\kern-.08em
    T\kern-.1667em\lower.7ex\hbox{E}\kern-.125emX}}
\begin{document}

\title{Are you a DePIN? \\
A Decision Tree to Classify \\ Decentralized Physical Infrastructure Networks
%\\ A Decision Tree for Classifying Decentralized and Token-Incentivized Infrastructure Systems\\
}

\author{\IEEEauthorblockN{1\textsuperscript{st} Michael S. Andrew}
\IEEEauthorblockA{\textit{Founder} \\
\textit{Who Loves Burrito?}\\
Portland, USA \\
w.l.burrito@wholovesburrito.com}
\and
\IEEEauthorblockN{2\textsuperscript{nd} Mark C. Ballandies}
\IEEEauthorblockA{\textit{Co-founder} \\
\textit{WiHi}\\
Zug, Switzerland \\
bcmark@protonmail.com}
}

\maketitle

\begin{abstract}

Decentralized physical infrastructure networks (DePINs) are an emerging vertical within "Web3" replacing the traditional method that physical infrastructures are constructed. Yet, the boundaries between DePIN and traditional method of building crowd-sourced infrastructures such as citizen science initiatives or other Web3 verticals are not always so clear cut. In this work, we systematically analyze the differences between DePIN and other Web2 and Web3 verticals. For this, the study proposes a novel decision tree for classifying systems as DePIN. This tree is informed by prior studies and differentiates DePIN from related concepts using criteria such as the presence of a three-sided market, token-based incentives for supply, and the requirement for physical asset placement in those systems.

The paper demonstrates the application of the decision tree to various blockchain systems, including Helium and Bitcoin, showcasing its practical utility in differentiating DePIN systems. 

This research offers significant contributions towards establishing a more objective and systematic approach to identifying and categorizing DePIN systems. It lays the groundwork for creating a comprehensive and unbiased database of DePIN systems, which will inform future research and development within this emerging sector.

\end{abstract}


\section{Introduction}

Distributed infrastructure has a long history, with systems like distributed.net and SETI@home pioneering distributed computing in the late 1990s. SETI@home, launched in 1999, allowed volunteers to contribute idle computer processing power to analyze radio signals for signs of extraterrestrial intelligence, making it one of the earliest and most successful examples of distributed infrastructure for scientific research~\cite{korpela2001seti}. Grid computing emerged around the same time as a means to harness the collective power of geographically dispersed computers for large-scale tasks. Systems like the Globus Toolkit \cite{foster2006globus} and the European DataGrid \cite{segal2000grid} were instrumental in developing distributed computing infrastructures. Peer-to-Peer (P2P) sharing networks like Napster also relied on distributed infrastructure, enabling users to share files directly without a central server. 
More recently, blockchain technology, popularized by cryptocurrencies like Bitcoin, has utilized distributed infrastructure to create secure and tamper-resistant networks.

Within blockchain and the web3, systems like Helium \cite{jagtap2021federated} pioneered a new wave of three-sided markets \cite{ballandies2023threesided} that bootstrap the supply of infrastructure, i.e. deployment of specialize hardware, incentivized by payment in blockchain tokens \cite{ballandies2023taxonomy}. This combination of cryptocurrency and distributed infrastructure has birthed a new sector in the blockchain realm, commonly noted as Decentralized Physical Infrastructure Networks or 'DePIN'.
Prior to the use of the term 'DePIN', similar 'real-world' blockchain systems were referred to by several different terms such as MachineFi, Proof of Useful Work, Token-Incentivized Phsyical Infrastructure Networks (TIPIN), Economy of Things, etc.   Based on an informal Twitter (now X) poll, the term DePIN was then adopted by analytics firm Messari \cite{messari2022DePINname}.
Since the publishing of this poll, the term has been widely adopted as a 'catch-all' for systems based on blockchain incentives and a replacement of Web2 services. Although the term is widely adopted, a standard definition has not be agreed upon with any consensus. This has led to a a proliferation of use of the term in marketing materials or application to systems such as Bitcoin and other traditional Proof of Work cryptocurrency mining. 
Even Messari has equivocated on the original application of the term by proposing two sub-sectors, Physical Resource Networks (PRN) and Digital Resource Networks (DRN) \cite{kassab2023navigating}, whereas it would appear the use of 'Physical' in the PRN term is redundant and possibly contradictory, and an admission that DePIN may not have been the best choice as an umbrella term.  Others have proposed a similar bisected taxonomy based primarily on the idea of fungibility of the resources. The designations of DePIN and DeReN (Decentralized Resource Networks) were coined to describe these two sub-sectors \cite{nystrom2023deren}.
A more recent Messari research report has claimed there are greater than 650 'DePIN' systems \cite{gala2023state}. However, this report has offered no definitive definition of 'DePIN'.  A review of some of the referenced systems indicates inclusion of blockchains such as IoTex and Peaq that, although designed and marketed to DePIN systems, are not exclusive to this sector.  A system can choose to build on these Layer 1 (L1) blockchains, regardless of sector designation.  Other systems , such as versasity\footnote{https://verasity.io/, last access: 2024-03-05} or Braitrust\footnote{https://www.usebraintrust.com/, last access: 2024-03-5}, do not require the contribution of physical elements. Thus, in absence of specific criteria for inclusion, we are left with the "I know it when I see it" explanation famously used in the US Supreme Court Case  Roth v. United States which is often criticized as being arbitrary \cite{gewirtz1995know}.

 This research seeks to create an unbiased, objective, and repeatable 'litmus test' to determine if systems are eligible for listing in a database that strives to be a neutral clearinghouse of systems within the DePIN sector. This clearinghouse strives to be uncluttered by marketing claims and focus on providing clear information for those who may be interested in a specific system type. In particular, in order to create a neutral database of systems, objective criteria for system inclusion are required. Absent of this, subjective decisions regarding inclusion may be considered as arbitrary and potentially biased.

 This work contributes such criteria in form of a decision tree, illustrates its application in case studies to blockchain systems such as Bitcoin, Helium and Akash, and discusses its implication on terminology and highlights limitations. 

\begin{figure*}[t]
    \centering
    \includegraphics[width=0.9\textwidth]{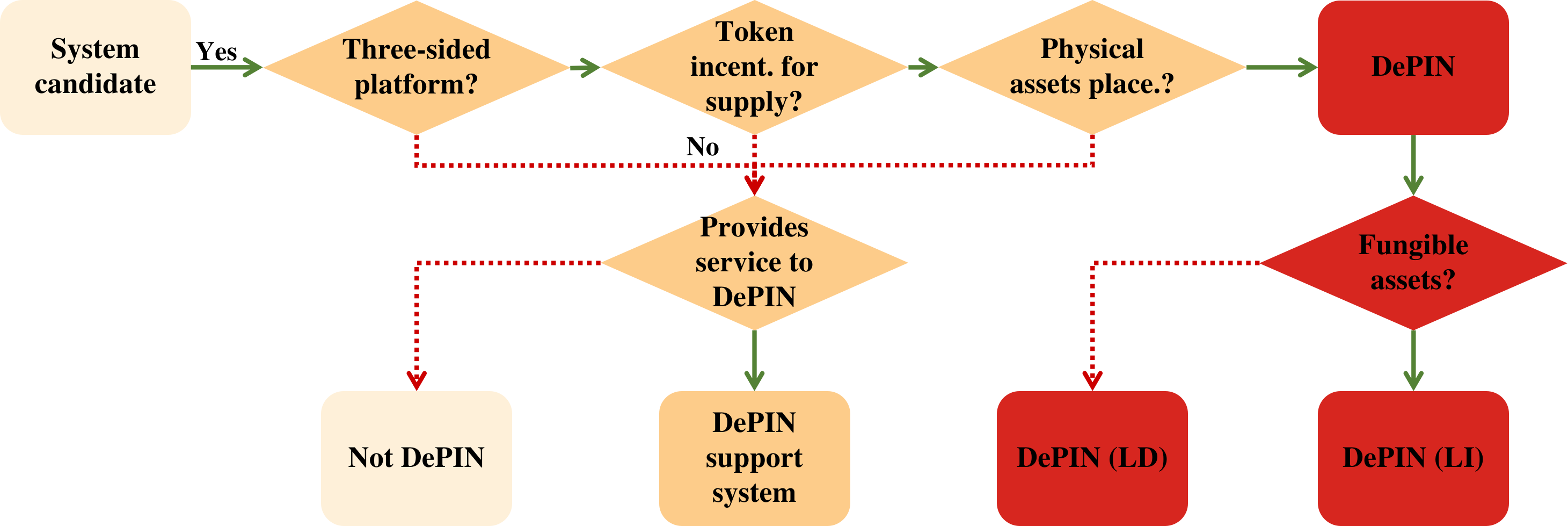}
    \caption{Proposed decision Tree for DePIN identification.}
    \label{fig:decision_tree}
\end{figure*}

\section{Related Work}
In previous blockchain/Web3 research, a decision tree has been used to distinguish between previously undefined or not clearly delineated terms such as DeFi/CeFi \cite{qin2021cefi}.

In the context of DePIN, several definitions and contributions have been presented from which such a distinction possibly could be derived. A previous study established a useful taxonomy to describe the various components and attributes of DePIN systems \cite{ballandies2023taxonomy}. This taxonomy does not seek to provide a definitive definition of DePIN; rather, it suggests that "whenever the incentivized action in a system involves the placement or contribution of physical infrastructure elements, such as a camera or storage drives, it can be defined as a DePIN system."
This aligns with the proposal to define DePIN as networks using blockchain-based token incentives to build real-world physical infrastructure networks \cite{gala2023state,peaq2024def} which further specifies the token-incentivized nature of the DePIN supply-side. This token incentivization has actually been stated as a differentiation criteria between DePIN and Web2 participatory sensing systems \cite{chiu2024tokenparticipatory}. 

Furthermore, it has been  proposed to differentiate between DePIN and DeREN systems, distinguishing between the fungible and non-fungible nature of assets comprising an infrastructure network \cite{nystrom2023deren}.

However, to our knowledge, clear criteria for classifying a system as DePIN that combine and extend these different definitions have not yet been established. This work combines and extends the above definitions of DePIN systems into a decision tree for DePIN classification.

\section{Methodology}
The decision tree in this study was derived through convenience sampling, by analyzing systems and suggestions crowd-sourced from a DePIN database website\footnote{The name is anonymized and will be published in final publication}. The DePIN community suggested 140 systems as DePIN or DePIN support systems to the authors. This dataset was extended by systems from other Web3 and Web2 verticals. Utilizing this training set, the authors qualitatively formulated criteria for unambiguous classification of systems as DePIN, subsequently constructing a decision tree from it. Each path from root to leaf in this tree represents a specific classification rule. This method represents an 'empirical-to-conceptual' iteration in the recognized taxonomy development methodology \cite{nickerson2013method}. A similar flow-like decision tree was previously qualitatively developed and employed to distinguish between centralized and decentralized finance systems \cite{qin2021cefi}.

\section{Decision Tree}
\label{sec:tree}
  
Figure \ref{fig:decision_tree} illustrates the decision tree.  
A system candidate has to fulfill the following three requirements in order to be considered a DePIN:

\textit{Three-sided platform}: 
A side of a platform is defined as an interface to a distinct user-group.
A market is said to be created on this interface when market mechanism are utilized to characterize the exchange of goods and services at this interface.
1-sided platforms for infrastructure are traditionally, for instance, created by utility companies by building an infrastructure (cell towers, power grid) on which a service is build (mobile or energy plans) that is sold to customers, i.e. to which their platform is faced. As a service is offered in exchange for fiat money, these one-sided platforms are one-sided markets.

Two-sided markets not only interface with customers, but also have an interface to a supply-side. For instance, the company Uber links infrastructure providers (car drivers) with users (passengers) via different interfaces.
Two-sided markets, in contrast to one-sided markets, make the suppliers revenue dependent on the platforms success with the demand side \cite{rysman2009economics}. %In the case of uber, market mechanisms are employed on both sides, thus creating a two-sided market.

DePIN systems can be characterized as three-sided platforms consisting of a supply, a service and a demand side~\cite{ballandies2023threesided} in which the supply-side is decentralized by creating a market that incentivizes a community to provide infrastructure resources. In particular, and similar to traditional two-sided markets like Uber, in a DePIN system the supply is provided by different actors than those maintaining the service of the system. This differentiates DePIN systems from L1 blockchain networks where the miner/ consensus participants directly provide the supply to users, thus removing the intermediary, hence exhibiting a two-sided platform layout. 

Traditional three-sided platforms like Uber often face the ‘cold-start’ problem \cite{eisenmann2006strategies} which illustrates that the supply first needs to be build in order for a service to be offered, but without a service it is hard/ costly to incentivize a supply to be build in the first place.  DePIN projects use the speculative value of tokens (see next criterion) to incentivize the supply side of the market. 

\textit{Token-based incentives for supply}: The supply side is incentivized to provide infrastructure to the network by earning blockchain-based tokens. These tokens can represent value within the network and can be used for various purposes such as accessing services, voting on governance issues, or trading on exchanges \cite{ballandies2023taxonomy}. This criterion is required to differentiate DePIN from Web2 citizen science/ participatory sensing initiatives \cite{chiu2024tokenparticipatory} or other traditional 3-sided platforms. 
DePIN projects can more easily incentivize infrastructure build-out (see next criterion) through these token-based compensation, as they are attractive to suppliers for their speculative value and potential for appreciation, even before market demand emerges.

\textit{Incentivized core economy action is physical assets placement}: The incentivized action in the core economy \cite{ballandies2023taxonomy} is the placement or contribution of physical infrastructure assets to the system.  These physical assets can take many forms such as sensors, cameras, storage drives, and processors. This criterion is required to differentiate DePIN from other Web3 verticals such as regenerative finance (ReFi) or decentralized science (DeSci). By separating the supply side from service side and incentivizing with blockchain tokens (see above criteria) a network of physical assets can potentially be deployed quickly and efficiently. 
    %\paragraph{Core economy contributes resources to applications beyond cryptocurrency} - This attribute refers to the ability to contribute resources, such as computing power, storage, or bandwidth, to applications beyond cryptocurrency transactions, thus eliminating traditional proof of work and proof of stake systems like Bitcoin and Ethereum.   

In case a system does not fulfill one of these criteria, but provides a service specific to a DePIN, it is considered a \textit{DePIN support system}.

The combination of the above criteria fuels the ‘Flywheel’ system that is considered the main engine of DePIN systems \cite{peaq2024def,gala2023state}. Systems that meet all of these attributes can effectively, and for the purpose of this research, be referred to as DePIN. 
Within this broader scope and based on the definition provided by \citet{nystrom2023deren} and following the reasoning given by \citet{gala2023state}, these systems can be further delineated in two sectors based on the following criteria:

\textit{Fungible assets}: Fungibility illustrates, in this context, the interchangeability of the physical assets that are placed in the system. In case a placed physical asset can be simply replaced by the contribution of another physical asset in the system, then the system is referred to as DePIN-LI (location independent). In case assets are not interchangeable, the system is referred to as DePIN-LD (location dependent). For example, in a decentralized storage system such as Filecoin\footnote{https://filecoin.io/, last access: 2024-03-05}, the specific location of the storage drive is not relevant, and if that storage goes offline, it can be replaced with another storage drive in any location with no real loss of utility.  On the other hand, we can examine a system like DIMO\footnote{https://dimo.zone/, last access: 2024-03-05}. DIMO is specific to a vehicle and removal of that device would result in loss of that specific data from the system.

% \begin{itemize}
%     \item Contributed physical elements are location dependant - This attribute means the physical elements contributed by the supply side are "location dependent". For example, a network of sensors providing real-time environmental data would be location dependent, as the sensors need to be physically located in specific places to collect relevant data. 
% \end{itemize}
% For the purposed of the attributes above, the terms 'core economy' and 'location dependant' are specifically defined as:
% \begin{itemize}
%     \item Core Economy - 
%     \item Location  -  The term 'Location' in this context, can be considered a physical geographic location (an address, a latitude/longitude/elevation, etc.), but may also be a place within a physical element (e.g. inside a car), or a proxy for location (e.g. residential IP address). 
% \end{itemize}
% Application of the attributes above lead us to propose three sub-sectors referred to the following and defined by the decision tree in Figure 1. 
% \begin{itemize}
%     \item DePIN-LD (Location Dependant)
%     \item DeReN-LI (Location Independent)

% \end{itemize}

% [add decision tree]

\section{Discussion}
\subsection{Case Studies}
Table \ref{tab:classification} illustrates a selection of systems in alphabetical order that could potentially fall under the term DePIN: 

% \usepackage{multirow}
\begin{table}[]
\caption{Classification of systems based on the decision tree illustrated in Figure \ref{fig:decision_tree}.}
\label{tab:classification}
\begin{tabular}{lllllll} \hline
\multirow{2}{*}{\textbf{System}} & \multicolumn{5}{c}{\textbf{Criteria}}                                                                                                                                                                                                                                                                                                                         & \multicolumn{1}{c}{\multirow{2}{*}{\textbf{Class.}}} \\
                                 & \textit{\begin{tabular}[c]{@{}l@{}}Three-\\sided\\ platform\end{tabular}} & \textit{\begin{tabular}[c]{@{}l@{}}Tokens\\ for\\supply\end{tabular}} & \textit{\begin{tabular}[c]{@{}l@{}}Phy. \\ assets\\ place.\end{tabular}} & \textit{Fungible} & \textit{\begin{tabular}[c]{@{}l@{}}DePIN\\ Service\end{tabular}} & \multicolumn{1}{c}{}                                         \\
\hline 
\textit{Akash}                   & Yes                                                                      & Yes                                                                    & Yes                                                                     & Yes               &                                                                  & DePIN-LI                                                \\
\textit{Bitcoin}                 & No                                                                       & Yes                                                                    & Yes                                                                     & Yes               & No                                                               & Not DePIN                                            \\
\textit{CCC}                     & Yes                                                                      & No                                                                     & Yes                                                                     & No                & No                                                               & Not DePIN                                            \\
\textit{Data Lake}               & Yes                                                                      & Yes                                                                    & No                                                                      & No                & No                                                               & Not DePIN                                            \\
\textit{Grass}                   & Yes                                                                      & Yes                                                                    & Yes                                                                     & Yes               &                                                                  & DePIN-LI                                               \\
\textit{Helium}                  & Yes                                                                      & Yes                                                                    & Yes                                                                     & No                &                                                                  & DePIN-LD                                             \\
\textit{IoTeX}                   & No                                                                       & Yes                                                                    & Yes                                                                     & Yes               & Yes                                                              & Support                                              \\
\textit{Peaq}                    & No                                                                       & Yes                                                                    & Yes                                                                     & Yes               & Yes                                                              & Support                                              \\
\textit{WiHi}                    & Yes                                                                      & Yes                                                                    & Yes                                                                     & No                &                                                                  & DePIN-LD                                            
                                                                         \\
\hline
\end{tabular}
\end{table}

\textit{The Akash Network} is a decentralized cloud computing platform that aims to provide a more efficient and cost-effective alternative to traditional cloud providers (e.g. Amazon Web Services).  There are several other decentralized compute and storage systems that offer a similar type of supply to the market for compute services and the Akash assessment can serve as a proxy for these other similar services. Akash creates a three-sided market place between suppliers and users of compute by providing a marketplace for connecting the suppliers and users. Contribution of physical compute assets are incentivized with the blockchain-based AKT token. Because the compute assets are fungible, Akash is classified as a DePIN-LI. 

\textit{The Climate City Cup (CCC)} \cite{dapp2021finance} is a citizen science/ participatory sensing initiative to crowd-source air quality data in cities by incentiving a friendly competition among the participating parties. On the supply side are the sensors placed by citizens, the service side is provided by CCC which aggregates the data, and on the demand side are policy makers. Though the initiative motivates the placement of physical non-fungible sensors, it does not incentivize participation with blockchain-based tokens. In the literal sense, this would be a decentralized infrastructure network, which, however, as already mentioned, does not correspond to the purpose of the currently observed designation 'DePIN', which requires token-based incentives. 

\textit{Data Lake} is a decentralized science (DeSci) system that incentivizes the sharing of patient data with researchers. Though it uses blockchain-based tokens to crowd-source data (supply), it does not incentivize the placement of physical infrastructure, thus it is not considered a DePIN. 

\textit{Grass} markets itself as selling an unused resource (e.g. extra internet bandwidth). Essentially, the software uses your IP address and internet connection to 'scrape' the web on behalf of the market side buyers. On the supply side are users who provide a computing device with capability to run a web-browser.  Grass provides the service and sells to  businesses that are interested to use the web scraped data (e.g. using it to train AI). Contributed computing assets are fungible, thus making Grass a DePIN-LI system.  

\textit{Helium} started as a decentralized wireless network focused on IoT (Internet of Things).  Helium is now divided into two networks governed by separate sub-DAO.  In this case, we are examining the IOT network (with IOT token as utility) but the findings are generally the same for the Mobile network (MOBILE token). 
Not surprisingly, Helium IOT is an uncontroversial DePIN system.  As shown in Figure 1, it meets all the attributes to be a DePIN, and specifically a DePIN-LD.  
%Notably, Helium rarely uses the term DePIN in it's marketing materials, but is commonly held out as an example by others that are attempting to explain 'What is a DePIN?'. 

\textit{Bitcoin}
%It would be presumed to be uncontroversial to exclude bitcoin as a DePIN, however, there have been arguments to the contrary \cite{}. Projects like Chia (using a consensus mechanism called "proof of space and time.") have also been found in so called 'DePIN' lists. 
can certainly be considered physical (in terms of miners), decentralized, and token-incentivized (earn BTC). %However, the spirit of DePIN, is that the service is contributing to a demand side market a utility that goes beyond what essentially amounts to continuance of the blockchain (validating and securing the chain via hashpower).  Therefore our decision tree specifically excludes proof of work or other similar L1 chains as shown in Figure 2.   
However, the spirit of DePIN, is that the service within the three-sided market is provided by others than those contributing the supply. In particular, Bitcoin can be considered as a two-sided market where those contributing the supply make up the platform that is given as a service to customers.  
  
\textit{IoTeX and Peaq} are L1 blockchains purposely built to service DePIN\footnote{https://docs.iotex.io/, https://docs.peaq.network/docs/quick-start/what-is-peaq/; last access: 2024-03-05}.   What differentiates these from other L1s are the 'add-ons'.  For example peaq refers to these as "easy-to-use Modular DePIN Functions".   Despite these DePIN-centric extras, like Bitcoin, these blockchains can be considered a two-sided market where those contributing the supply make up the platform that is given as a service to Web3 customers, thus not meeting the first criteria. However, both explicitly offer DePIN-focused services and have DePINs built on them, which classifies them as DePIN support systems.

\textit{WiHi\footnote{www.wihi.cc, last access: 2024-03-05}} consists of a three-sided market that collects weather data via physical non-fungible weather stations on the supply side and applies machine learning to it via its expert community; both are offered to users as a service. The supply side is incentivized with a token, making WiHi a DePIN-LD.

%\subsection{Braintrust}

\subsection{Terminology and Criteria}
It seems the community has identified the need to differentiate between DePIN-LI and DePIN-LD \cite{nystrom2023deren,gala2023state} as done in this work. To make the terms more accessible, it could be beneficial to follow the suggestion of \citet{nystrom2023deren} and refer to the former as DeREN and to the latter as DePIN. Both categories could then be summarized in the umbrella term of \textit{Decentralized and Token-Incentivized Infrastructure Systems (DeTIS)}.
Moreover, within each of these two categories, further sub-categories can be defined such as compute and storage in DeREN or sensors and DeWi in DePIN. A potential route forward could be to do this categorisation via the \textit{Device} attribute in the previously introduced DePIN taxonomy \cite{ballandies2023taxonomy}. 

Also, the three-sided platform criteria could further be split into a criteria requiring that the demand side addresses non-DePIN/ non-Web3 users to highlight the real-world impact DePIN systems usually have.

In addition, one could loosen the requirement of token-based incentives and extend it to scenarios where the blockchain technology is used for coordination (e.g. via DAO governance). In particular, tokens are only one instance of this capacity of blockchains to coordinate human behavior. 

Moreover, we have the intuition that "Fungibility" is only a proxy for a broader concept potentially better described by "location specificity".   Some projects such as Grass noted above, or some dVPN systems (e.g. Mysterium) require a residential IP address which is a form of 'location'. For the system to work, the physical device contribution must be in a specific 'location', which perhaps could be more broadly defined (e.g. in a car, at a residential IP, at a specific geolocation, etc.) and may be more conceptually acceptable than the 'interchangeability' as defined under 'Fungibility".  
For instance, looking at the earlier example of DIMO, although the network is more valuable with additional vehicles, the value of any individual unique vehicle is hard to define. If one takes away a vehicle and interchanged it with another similar vehicle would there be not likely be a significant loss of utility.  Perhaps the defining attribute is that the device is 'located' inside of a unique vehicle. Also, considering Helium, it is not that one cannot replace a device with another nearby and have the network work normally. It is that there is a specific type of 'location' where it must be placed in order to contribute. 

\subsection{Limitations}
\label{sec:limitations}
Although the systems used for developing the decision tree were crowd-sourced, suggesting some objectivity, the authors' involvement in this process introduces potentially a bias in the selection and classification of systems as DePIN/DeTIS.  

Moreover, the decentralized nature of DePIN systems is currently not part of the criteria, potentially leaving room for systems to be included that are not decentralized when it comes to the access to contribute supply-side physical assets. In an initial version of the decision tree, this was considered by having the criteria of supply-side "Actor Permission" \cite{ballandies2023taxonomy} in the core economy to be required to be 'open'. Such a system was not found in the analyzed dataset\footnote{Some systems due to their immature state maintain waitlists for supply-side onboarding. Due to their expressed plan to open this in the future, we classified them as open.} and thus this criteria was removed, but could be re-added once required. 

Further, in this context, the service-side (e.g. Middleware attributes \cite{ballandies2023taxonomy}) and demand-side of a DePIN system should also be open in order for a DePIN to be decentralized, thus making it a three-sided market according to Section \ref{sec:tree}. 
Also, one can raise the question if a DePIN is decentralized in cases where, for instance, the governance of the network is centralized as it has been found to be the case in some DePIN systems \cite{ballandies2023taxonomy}. 
However, the current objective of the decision tree is to classify systems as DePIN based on prevailing definitions and uses of the term, rather than to theorize on the ideal characteristics of a DePIN.

\section{Outlook}
This work contributes a decision tree for identifying DePIN systems based on the three criteria of three-sided platform, token-based incentives and physical asset placement, illustrates its applicability to classify blockchain systems as DePIN and discusses its implications and limitations. This decision tree is seen as a work in progress which will be refined in iterative interactions with the community. In particular, a community-wide evaluation of the decision tree is envisioned to strengthen its usefulness. For instance, the three-sided platform nature of DePIN systems and the fungibility criteria could be further differentiated as discussed.

Such a decision tree is a necessary building block to create an objective dataset of DePIN systems that will inform and support further research in this emerging web3 vertical. In particular, such a dataset could be used to identify key design decisions in DePIN systems by analyzing design configurations of DePIN systems and thus help interested parties understand and innovate the ecosystem as it was performed for blockchain systems earlier \cite{spychiger2021unveiling,ballandies2022decrypting}. Also, the categorization into DeREN and DePIN could be extended into further subcategories that would eventually provide a comprehensible overview of the DePIN landscape. Finally, as identified in Section \ref{sec:limitations}, one could start elaborating how an ideal DePIN system should look like by incorporating decentralization as a necessary criteria for all components of a DePIN system.

\bibliographystyle{IEEEtranN}
\bibliography{depin}

% Generated by IEEEtranN.bst, version: 1.14 (2015/08/26)
\begin{thebibliography}{20}
\providecommand{\natexlab}[1]{#1}
\providecommand{\url}[1]{#1}
\csname url@samestyle\endcsname
\providecommand{\newblock}{\relax}
\providecommand{\bibinfo}[2]{#2}
\providecommand{\BIBentrySTDinterwordspacing}{\spaceskip=0pt\relax}
\providecommand{\BIBentryALTinterwordstretchfactor}{4}
\providecommand{\BIBentryALTinterwordspacing}{\spaceskip=\fontdimen2\font plus
\BIBentryALTinterwordstretchfactor\fontdimen3\font minus \fontdimen4\font\relax}
\providecommand{\BIBforeignlanguage}[2]{{%
\expandafter\ifx\csname l@#1\endcsname\relax
\typeout{** WARNING: IEEEtranN.bst: No hyphenation pattern has been}%
\typeout{** loaded for the language `#1'. Using the pattern for}%
\typeout{** the default language instead.}%
\else
\language=\csname l@#1\endcsname
\fi
#2}}
\providecommand{\BIBdecl}{\relax}
\BIBdecl

\bibitem[Korpela et~al.(2001)Korpela, Werthimer, Anderson, Cobb, and Leboisky]{korpela2001seti}
E.~Korpela, D.~Werthimer, D.~Anderson, J.~Cobb, and M.~Leboisky, ``Seti@ home-massively distributed computing for seti,'' \emph{Computing in science \& engineering}, vol.~3, no.~1, pp. 78--83, 2001.

\bibitem[Foster(2006)]{foster2006globus}
I.~Foster, ``Globus toolkit version 4: Software for service-oriented systems,'' \emph{Journal of computer science and technology}, vol.~21, pp. 513--520, 2006.

\bibitem[Segal et~al.(2000)Segal, Gagliardi, Robertson, and Carminati]{segal2000grid}
B.~Segal, F.~Gagliardi, L.~Robertson, and F.~Carminati, ``Grid computing: the european data grid project,'' 2000.

\bibitem[Jagtap et~al.(2021)Jagtap, Yen, Wu, Schulman, and Pannuto]{jagtap2021federated}
D.~Jagtap, A.~Yen, H.~Wu, A.~Schulman, and P.~Pannuto, ``Federated infrastructure: usage, patterns, and insights from" the people's network",'' in \emph{Proceedings of the 21st ACM Internet Measurement Conference}, 2021, pp. 22--36.

\bibitem[Ballandies(2023)]{ballandies2023threesided}
M.~C. Ballandies. (2023) Depins are three-sided markets — an evaluation guide for investors to access decentralized physical infrastructure networks.

\bibitem[Ballandies et~al.(2023)Ballandies, Wang, Chung Chee~Law, Yang, G{\"o}sken, and Andrew]{ballandies2023taxonomy}
M.~C. Ballandies, H.~Wang, A.~Chung Chee~Law, J.~C. Yang, C.~G{\"o}sken, and M.~Andrew, ``A taxonomy for blockchain-based decentralized physical infrastructure networks (depin),'' \emph{Center for Law \& Economics Working Paper Series}, vol.~3, 2023.

\bibitem[Messari(2022)]{messari2022DePINname}
\BIBentryALTinterwordspacing
Messari. (2022) Web3 physical infrastructure needs a name! [Online]. Available: \url{https://x.com/MessariCrypto/status/\-1588938954807869440?s=20}
\BIBentrySTDinterwordspacing

\bibitem[Kassab(2023)]{kassab2023navigating}
S.~Kassab. (2023) Navigating the depin domain.

\bibitem[Nystrom(2023)]{nystrom2023deren}
\BIBentryALTinterwordspacing
M.~Nystrom. (2023) Depin and deren: Toward a better classification of decentralized infrastructure networks. [Online]. Available: \url{https://variant.fund/articles/depin\--deren-toward-better-classification\--decentralized-infrastructure-networks/}
\BIBentrySTDinterwordspacing

\bibitem[Gala and Kassab(2024)]{gala2023state}
\BIBentryALTinterwordspacing
S.~Gala and S.~Kassab. (2024) State of depin 2023. [Online]. Available: \url{https://messari.io/report/state-of-depin\--2023}
\BIBentrySTDinterwordspacing

\bibitem[Gewirtz(1995)]{gewirtz1995know}
P.~Gewirtz, ``On i know it when i see it,'' \emph{Yale LJ}, vol. 105, p. 1023, 1995.

\bibitem[Qin et~al.(2021)Qin, Zhou, Afonin, Lazzaretti, and Gervais]{qin2021cefi}
K.~Qin, L.~Zhou, Y.~Afonin, L.~Lazzaretti, and A.~Gervais, ``Cefi vs. defi--comparing centralized to decentralized finance,'' \emph{arXiv preprint arXiv:2106.08157}, 2021.

\bibitem[peaq(2023)]{peaq2024def}
\BIBentryALTinterwordspacing
peaq. (2023) Depin: What are decentralized physical infrastructure networks? [Online]. Available: \url{https://www.peaq.network/blog/what-are-\-decentralized-physical-infrastructure\--networks-depin}
\BIBentrySTDinterwordspacing

\bibitem[Chiu et~al.(2024)Chiu, Mahajan, Ballandies, and Kalabi\'{c}]{chiu2024tokenparticipatory}
M.~T.~C. Chiu, S.~Mahajan, M.~C. Ballandies, and U.~V. Kalabi\'{c}, ``Depin: A framework for token-incentivized participatory sensing,'' 2024.

\bibitem[Nickerson et~al.(2013)Nickerson, Varshney, and Muntermann]{nickerson2013method}
R.~C. Nickerson, U.~Varshney, and J.~Muntermann, ``A method for taxonomy development and its application in information systems,'' \emph{European Journal of Information Systems}, vol.~22, no.~3, pp. 336--359, 2013.

\bibitem[Rysman(2009)]{rysman2009economics}
M.~Rysman, ``The economics of two-sided markets,'' \emph{Journal of economic perspectives}, vol.~23, no.~3, pp. 125--143, 2009.

\bibitem[Eisenmann et~al.(2006)Eisenmann, Parker, Van~Alstyne, et~al.]{eisenmann2006strategies}
T.~Eisenmann, G.~Parker, M.~W. Van~Alstyne \emph{et~al.}, ``Strategies for two-sided markets,'' \emph{Harvard business review}, vol.~84, no.~10, p.~92, 2006.

\bibitem[Dapp et~al.(2021)Dapp, Helbing, and Klauser]{dapp2021finance}
M.~M. Dapp, D.~Helbing, and S.~Klauser, \emph{Finance 4.0-Towards a Socio-Ecological Finance System: A Participatory Framework to Promote Sustainability}.\hskip 1em plus 0.5em minus 0.4em\relax Springer Nature, 2021.

\bibitem[Spychiger et~al.(2021)Spychiger, Tasca, and Tessone]{spychiger2021unveiling}
F.~Spychiger, P.~Tasca, and C.~J. Tessone, ``Unveiling the importance and evolution of design components through the “tree of blockchain”,'' \emph{Frontiers in Blockchain}, vol.~3, p. 613476, 2021.

\bibitem[Ballandies et~al.(2022)Ballandies, Dapp, and Pournaras]{ballandies2022decrypting}
M.~C. Ballandies, M.~M. Dapp, and E.~Pournaras, ``Decrypting distributed ledger design—taxonomy, classification and blockchain community evaluation,'' \emph{Cluster computing}, vol.~25, no.~3, pp. 1817--1838, 2022.

\end{thebibliography}

% \bibitem{b1} G. Eason, B. Noble, and I. N. Sneddon, ``On certain integrals of Lipschitz-Hankel type involving products of Bessel functions,'' Phil. Trans. Roy. Soc. London, vol. A247, pp. 529--551, April 1955.
% \bibitem{b2} J. Clerk Maxwell, A Treatise on Electricity and Magnetism, 3rd ed., vol. 2. Oxford: Clarendon, 1892, pp.68--73.
% \bibitem{b3} I. S. Jacobs and C. P. Bean, ``Fine particles, thin films and exchange anisotropy,'' in Magnetism, vol. III, G. T. Rado and H. Suhl, Eds. New York: Academic, 1963, pp. 271--350.
% \bibitem{b4} K. Elissa, ``Title of paper if known,'' unpublished.
% \bibitem{b5} R. Nicole, ``Title of paper with only first word capitalized,'' J. Name Stand. Abbrev., in press.
% \bibitem{b6} Y. Yorozu, M. Hirano, K. Oka, and Y. Tagawa, ``Electron spectroscopy studies on magneto-optical media and plastic substrate interface,'' IEEE Transl. J. Magn. Japan, vol. 2, pp. 740--741, August 1987 [Digests 9th Annual Conf. Magnetics Japan, p. 301, 1982].
% \bibitem{b7} M. Young, The Technical Writer's Handbook. Mill Valley, CA: University Science, 1989.
% \end{thebibliography}
\vspace{12pt}
\color{red}

\end{document}